# Silicon nitride gate dielectrics and bandgap engineering in graphene layers

Wenjuan Zhu[*], Deborah Neumayer, Vasili Perebeinos and Phaedon Avouris[**]
IBM Thomas J. Watson Research Center, Yorktown Heights, NY 10598, USA

**Abstract**
**We show that silicon nitride can provide uniform coverage of graphene in field-effect transistors while preserving the channel mobility. This insulator allowed us to study the maximum channel resistance at the Dirac (neutrality) point as a function of the strength of a perpendicular electric field in top-gated devices with different numbers of graphene layers. Using a simple model to account for surface potential variations (electron-hole puddles) near the Dirac point we estimate the field-induced band-gap or band-overlap in the different layers.**

Graphene as a two dimensional material shows remarkable electrical [1,2] mechanical [3], and optical properties [4]. Its high intrinsic carrier mobilities [5,6] make it a very promising material for electronic devices, particularly for analogue high frequency devices[7]. One important challenge in the use of graphene in devices is the difficulty of depositing gate dielectrics on it due to its hydrophobic nature. Given that a perfect graphene surface is chemically inert, direct growth of high dielectric constant gate insulators by atomic layer epitaxy (ALD) on a clean graphite surface usually leads to discontinuous films, where the dielectrics preferably grow on steps or defect sites which serve as nucleation centers [8, 9, 10]. A number of surface pretreatments, such as exposure to $NO_2$ [11], PTCA (carboxylate-terminated perylene) [10], ozone [8] or a seed layer such as Al [12] have been investigated; however these pretreatments or nucleation seed layers usually severely degrade the graphene channel mobility. Deposition of silicon oxide by evaporation has also been tested and was found to decrease the channel mobility severely (up to 85% decrease of the initial mobility) [13]. A low dielectric constant polymer, NFC, was shown to effectively cover graphene as a seed layer for subsequent ALD insulator deposition and to preserve the good graphene transport properties [14]. Nevertheless, for technological applications it would be desirable to avoid the liquid processing required and also increase both the thermal stability and the dielectric constant of the gate insulator.

In this work, we show that silicon nitride can be directly deposited on graphene without the help of a seed layer to obtain excellent coverage. The channel mobility remains largely intact after silicon nitride deposition. Furthermore, using top-gated devices with the silicon nitride gate dielectric, we were able to apply high electric fields and study the electrical band-gap opening or band-overlap generated in devices with different numbers of graphene layers as a function of an applied vertical electric field. The effect of an E-field on the band structure of bi-layer graphene is a subject of considerable current interest [15,16,17,18,19,20]. Studies of optical band-gap opening in bilayer

Corresponding authors: [*] wenjuan@us.ibm.com, [**] avouris@us.ibm.com

have been published [21,22,23] and the relation between optical and electrical band-gaps has been discussed [24].

The graphene layers were deposited through mechanical exfoliation of Kish graphite on a 90nm SiO$_2$ film grown on a silicon substrate. The resistivity of the silicon substrate was $10^3$~$10^4$ ohm-cm. The number of layers deposited was determined by measuring the changes in the reflectance of green light [25,26]. Hall-bar structures were fabricated using oxygen plasma. The silicon nitride gate dielectrics were deposited by plasma enhanced chemical vapor deposition (PECVD) in a 200mm AMAT DXZ chamber. The films were deposited at 400°C using SiH$_4$, NH$_3$, and N$_2$ with a HFRF plasma of 40 watts. The electrodes were made of Ti/Pd/Au. The gate lengths (the distance between two voltage sensing terminals) used in our devices were 2μm and the gate widths were 0.5μm. The electrical measurements were performed in high vacuum ($10^{-6}$~$10^{-8}$ Torr).

Figure 1 (a) shows the SEM image of 10nm PECVD silicon nitride film deposited on highly ordered pyrolytic graphite (HOPG). We see that the silicon nitride film gives a very uniform coverage on graphite with no visible pin-holes. For comparison, the SEM image of CVD silicon oxide on HOPG is shown in Figure 1 (b). The silicon oxide is deposited by thermal CVD using tetraethyl orthosilicate (TEOS) and ozone at 400°C. The film is now discontinuous with large pin-holes. These results demonstrate that PECVD silicon nitride adheres much better to the graphite surface than silicon dioxide deposited by the thermal process. One possibility for this is a favorable initial interaction of NH$_3$ with the graphene surface, which forms effectively a seed layer. Another possibility is related to the PECVD process itself. PECVD depositions are not surface controlled reactions, but involve gas phase reactions with the decomposition of the precursors initiated in the plasma and are thus less sensitive to the nature of the substrate. In the silicon oxide CVD process, due to the involvement of an oxidizing ambient (ozone), plasma use must be avoided due to the etching of graphene. However, in the silicon nitride CVD process, an inert ambient is used (N$_2$ and NH$_3$), and thus a plasma can be applied. The PECVD silicon nitirde process was engineered to be a low density plasma to minimize surface damage to graphene. The low density plasma was obtained by striking and sustaining plasma with only 40 watts. A high pressure of 8 torr and a large N$_2$ flow of 10000 sccm were utilized to reduce the mean free path of radicals in the plasma by maximizing collisions within the plasma and not with the film surface. The resultant plasma is relatively cold and mild. Consistent with the low plasma density is the low silicon nitride growth rate of 5Å/s. All of these factors contribute to minimize damage to the graphene and to ensure continuous growth of silicon nitirde over the graphene.

Our capacitance and ellipsometry analysis yield a dielectric constant of $\varepsilon = 6.6$ for the resulting silicon nitride films, which is significantly higher than the dielectric constant of silicon dioxide ($\varepsilon = 3.9$) and that of NFC ($\varepsilon = 2.4$) [14]. We have also measured the breakdown field of the silicon nitride films by fabricating MOS capacitors and find a high average breakdown field of 11.5MV/cm for capacitors with an area of ~1300μm$^2$, as shown in Figure 2. The MOS structure is 1nm Ti/20nm Pd/40nm Au electrode with 20.2nm silicon nitride gate dielectric on highly doped p-type silicon substrate. An additional advantage of silicon nitride insulators over silicon oxide for graphene devices is their higher surface polar optical phonon frequency ~110meV [27, 28, 29] vs. ~56meV [30] for silicon oxide, which should decrease the importance of remote inelastic phonon scattering in the graphene channel [31,32].

Figure 3 shows the results of electrical measurements on the devices. First the drain current versus gate voltage of the back-gated graphene transistors before and after silicon nitride deposition are shown. The field effect mobilities of the graphene channel before and after nitride deposition are given in the insets. The field effect mobility is calculated from: $\mu_{FE} = \frac{1}{C_{BG}} \frac{\partial \sigma}{\partial V_{BG}}$, where $\sigma$ is the four probe conductivity, $C_{BG}$ is the back-gate capacitance and $V_{BG}$ is the back-gate voltage. The mobility after nitride deposition is comparable or even higher than the one before nitride deposition for bi-, tri- and multi-layer (4~5 layers) graphene. In part this can be explained by the increased dielectric constant (silicon nitride vs. vacuum) which screens charged impurity scattering from the underlying $SiO_2$. On the other hand, the mobility is slightly lower for mono-layer graphene indicating that another type of scattering, likely short-range, is increased. The modification of the short-range scattering in bi-, tri- and multi-layers should have a smaller effect on the overall mobility than in a single layer, because the outer layer would be mostly affected by it. This is also suggested by the larger mobility increase observed in the tri-layer than in the bi-layer, as seen in the insets of Figure 3b and 3c.

Using silicon nitride top-gated graphene devices, we studied the changes in the channel resistance and the induced band-gap or band-overlap as a function of the strength of a perpendicular electric field. Figures 4(a) to (d) show the four-point channel resistance as a function of top-gate voltage at different back-gate voltages for the different devices. We see that the maximum resistance, i.e. resistance at the Dirac/neutrality point, is nearly unchanged for mono-layer graphene, while it changes significantly for bi-, tri- and multi-layer graphenes. The maximum channel resistance as a function of the average displacement field $D_{Ave}$ is shown in Figure 5. Here the average displacement field is defined as [15]: $D_{Ave} = (D_b + D_t)/2$, where $D_t$ is the top-gate displacement field: $D_t = -\varepsilon_t(V_{TG} - V_{TG}^0)/d_t$, and $D_b$ is the back-gate displacement field: $D_b = +\varepsilon_b(V_{BG} - V_{BG}^0)/d_b$. Here $\varepsilon$ and $d$ are the dielectric constant and thickness of the dielectric layer and $V^0$ is the Dirac offset voltage due to the initial environmentally induced carrier doping. The dielectric constants are: $\varepsilon_b = 3.9$ for the back-gate silicon oxide and $\varepsilon_t = 6.6$ for the top-gate silicon nitride. At the Dirac or neutrality points, $D_b = D_t$, so the average displacement field can be written as $D_{Ave} = D_b = \varepsilon_b(V_{BG} - V_{BG}^0)/d_b$. We see that as the average displacement field $D_{Ave}$ increases, the maximum resistance increases for bi- and tri-layer graphene, while it decreases for multi-layer graphene (4~5 layers).

The above observed trend can be accounted by the band-gap or overlap changes induced by the increasing vertical field. However, before proceeding with the evaluation of the values of the induced band-gaps/overlaps we need to consider the variation of the surface potential landscape near the Dirac/neutrality point. Electrons and holes form puddles [33, 34, 35]. These puddles can then strongly influence the transport physics at low carrier densities. In our analysis we use the simple model [36] illustrated in Figure 6(a). Here the surface electrostatic potential is described as a step function. We define $\Delta$ as the half height of the peak-to-peak variation in electrostatic potential and $\Phi$ as half of the band gap/overlap: $\Phi = (E_c - E_v)/2$, where $E_c$ is the conduction band energy and $E_v$ is the valence band energy. A value of $\Phi > 0$, implies a band-gap opening, while $\Phi < 0$ implies a band-overlap. When $D_{Ave} = 0$, (i.e. there is no induced band-gap/overlap), the carrier density at the neutrality point can be expressed as follows for bi-, tri- or multi-layer graphene [36]:

$$n_{Dirac}^0 = \frac{2m^*}{\pi\hbar^2} k_B T \ln[(1 + e^{\Delta/k_B T})(1 + e^{-\Delta/k_B T})] \tag{1}$$

where the carrier density at Dirac point $n_{Dirac}^0$ can be extracted by fitting the conductivity $\sigma$ as a function of the top-gate voltage $V_{TG}$, based on the equation $\sigma = \mu\sqrt{(en_{Dirac}^0)^2 + [C_{TG}(V_{TG} - V_{TG}^0)]^2}$, where $C_{TG}$ is the top-gate capacitance. When the average displacement field is not zero, i.e. when there is a finite induced band-gap/overlap, the carrier density at the neutrality point can be expressed as:

$$n_{Dirac} = \frac{2m^*}{\pi\hbar^2} k_B T \ln[(1 + e^{(\Delta-\Phi)/k_B T})(1 + e^{-(\Delta+\Phi)/k_B T})] \tag{2}$$

If the mobility is independent of the average displacement field, then the resistance ratio $R_{Dirac}^0 / R_{Dirac}$ can be written as:

$$\frac{R_{Dirac}^0}{R_{Dirac}} = \frac{n_{Dirac}}{n_{Dirac}^0} \tag{3}$$

Combining equations (1-3), we obtain:

$$\frac{R_{Dirac}^0}{R_{Dirac}} = \frac{\ln[(1 + e^{(\Delta-\Phi)/k_B T})(1 + e^{-(\Delta+\Phi)/k_B T})]}{\ln[(1 + e^{\Delta/k_B T})(1 + e^{-\Delta/k_B T})]} \tag{4}$$

When $\Delta = 0$, i.e. with no surface electrostatic potential variation, equation (4) can be further simplified to:

$$\frac{R_{Dirac}^0}{R_{Dirac}} = \frac{\ln(1 + e^{-\Phi/k_B T})}{\ln 2} \tag{5}$$

From equations (4) and (5), we can now extract $\Phi$ at the different back-gate voltages or average displacement fields with and without considering the surface electrostatic potential, respectively. The insets in Figure 5 show the extracted band-gap/overlap $2\Phi$ as a function of the average displacement field. The breaking of symmetry by the vertical electric field in bilayer is expected to open a band-gap [19,20]. Indeed, the obtained positive $2\Phi$ indicates a band-gap opening and we obtain a gap of about 50meV at a field of 1.3Vnm$^{-1}$. This band-gap value is very similar to the electrical gap measured at the same field by Xia et al. [24] utilizing a combination of a polymer seed layer and a high-$\varepsilon$ ALD insulator and close to the value predicted by theoretical calculations[15,19]. In the case of the tri-layer device, $2\Phi$ is also positive indicating a band-gap opening. For an ABA tri-layer, theory predicts [37,38] that the field would induce a band-overlap and a band-overlap was reported before in a tri-layer device [26]. The fact that we observe a small gap instead suggests that the layer stacking in our device is not ABA. Either the layer stacking is ABC where a gap is expected, or most likely, the third layer is disoriented so that the system acts as a bilayer with the third, largely uncoupled, layer partially screening the electric field.

In multi-layer graphene, $2\Phi$ is negative, indicating the presence of band-overlap. For ABA graphite-like multilayers, the field is predicted to lead to increased band-overlap [37,38] in agreement with our experimental findings. In general, however, there is a possibility that the graphene layers in a multi-layer structure are randomly stacked. Such random stacking can be obtained, for example, in graphene grown by thermal decomposition of SiC. In this case, application of an electric field would lead to the displacement of the Dirac points of the different layers while preserving the linear band-dispersion, as was theoretically predicted in the case of bi-layer [39]. Illustrations of the bandstructures for disoriented multi-layers with and without a perpendicular displacement field are shown in Figures 6(f) and (g). In this case the resistance ratio should be given by:

$$\frac{R^0_{Dirac}}{R_{Dirac}} = \frac{\pi^2/3 + (\Delta/k_BT)^2 + (\Phi/k_BT)^2 + Li_2(-e^{(\Delta+\Phi)/k_BT}) + Li_2(-e^{-(\Delta-\Phi)/k_BT})}{\pi^2/3 + (\Delta/k_BT)^2 + Li_2(-e^{\Delta/k_BT}) + Li_2(-e^{-\Delta/k_BT})} \quad (6)$$

where $Li_n(x) = \sum_{k=1}^{\infty} \frac{x^k}{k^n}$ is the polylogarithm function. When $\Delta = 0$, i.e. with no surface electrostatic potential variation, equation (6) can be simplified to:

$$\frac{R^0_{Dirac}}{R_{Dirac}} = \frac{\pi^2/3 + (\Phi/k_BT)^2 + 2Li_2(-e^{\Phi/k_BT})}{\pi^2/6} \quad (7)$$

The values of the band-overlap extracted based on equations (6) and (7) are also plotted in the inset of Figure 5(d). We can see that the band-overlap extracted by these two methods, i.e. parabolic band overlap due to coupled layers, or linear band shifting from disoriented single layer stacks, are actually very similar.

We also note that when $\Phi >> k_BT$, equation (5) can be further simplified to: $\frac{R^0_{Dirac}}{R_{Dirac}} = \frac{1}{\ln 2} e^{-\Phi/k_BT}$. This result is consistent with the commonly used model based on thermionic emission of carriers over the metal-graphene Schottky barriers [40, 21] for $\Phi >> k_BT$.

In conclusion, we found that silicon nitride can provide excellent coverage of graphene in field-effect transistors while preserving its good carrier mobilities, without the need of a seed layer. Moreover, the silicon nitride film has the advantage of higher dielectric constant and higher surface polar optical phonon energy (i.e. less remote phonon scattering in the graphene channel) compared to silicon oxide. The breakdown strength in silicon nitride is high as well. The effect of a perpendicular electric field on the band-structure of different numbers of graphene layers used as channels of the transistor was also studied and the induced band-gap or band-overlap was obtained accounting for the effects of the variation of the surface potential near the Dirac/neutrality point.

**Acknowledgements**
We would like to thank B. Ek, J. Bucchignano, G. P. Wright and V. K. Lam for their contributions to device fabrication. We would also like to thank K. A. Jenkins for his help on characterizations. We are also grateful to F. Xia, H.-Y. Chiu, M. Freitag, Y.-M. Lin, D. Farmer and C.Y. Sung for their insightful discussions. This work is supported by DARPA under contract FA8650-08-C-7838 through the CERA program.

# I. Figure Captions

Figure 1. SEM images of (a) silicon nitride and (b) silicon oxide films deposited on HOPG.

Figure 2. Cumulative failure and histogram of the breakdown field for Au/Pd/Ti/$Si_3N_4$/Si MOS capacitors.

Figure 3. Conductivity before and after 10nm silicon nitride gate dielectrics deposition on (a) mono-layer, (b) bi-layer, (c) tri-layer and (d) multi-layer graphene (4~5 layers) respectively. The inset shows the field effect mobility before and after silicon nitride deposition on (a) mono-layer, (b) bi-layer, (c) tri-layer and (d) multi-layer graphene (4~5 layers) respectively.

Figure 4. Four point resistance as a function of top-gate voltage at various back-gate voltages for (a) mono-layer, (b) bi-layer, (c) tri-layer and (d) multi-layer graphene with 10nm silicon nitride gate dielectrics.

Figure 5. Maximum channel resistance as a function of the average displacement field for (a) mono-layer, (b) bi-layer, (c) tri-layer and (d) multi-layer graphene devices with 10nm silicon nitride gate dielectrics. The symbols are the experimental measurements and the lines are for guiding the eye. The insets in (b) and (c) show the band-gap, 2Φ, as a function of the average displacement field with and without considering the surface electrostatic potential Δ for bi-layer and tri-layers, respectively. The inset in (d) shows the 2Φ (band overlap) as a function of the average displacement field using the parabolic band model described by equations (4) and (5) and the linear band model using equations (6) and (7), with and without considering the surface electrostatic potential Δ.

Figure 6. (a) Illustration of the spatial inhomogeneity of the electrostatic potential, the band-gap (or overlap) and the model used in the analysis of the band gap/overlap for bi-layer, tri-layer and multi-layer graphene. ($\Phi > 0$ represents band-gap opening, while $\Phi < 0$ represents band-overlap). (b) and (c) illustrate the band gap opening under a vertical displacement field $D_{Ave}$ in bi/tri-layer graphene. (d) to (g) illustrate the two possible scenarios in band overlap generation under a vertical displacement field $D_{Ave}$ in multi-layer graphene: (e) parabolic band overlap and (g) linear band shifting.

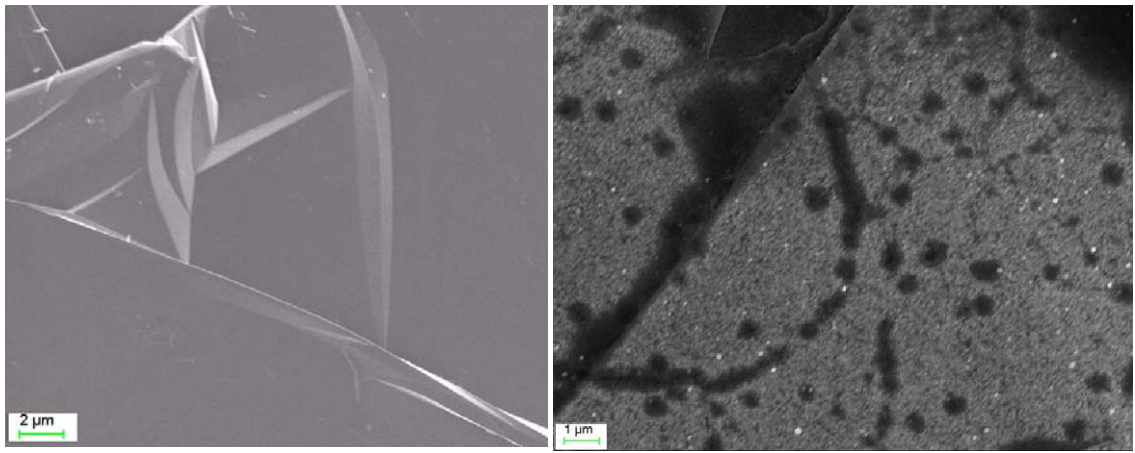

(a)                                            (b)

Figure 1.

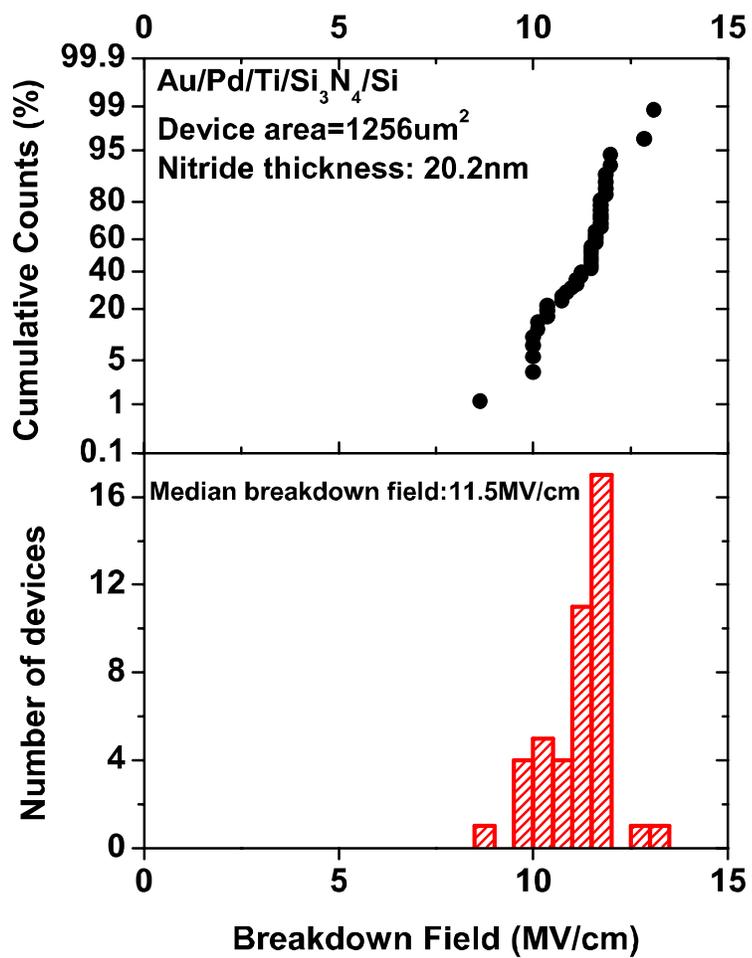

Figure 2

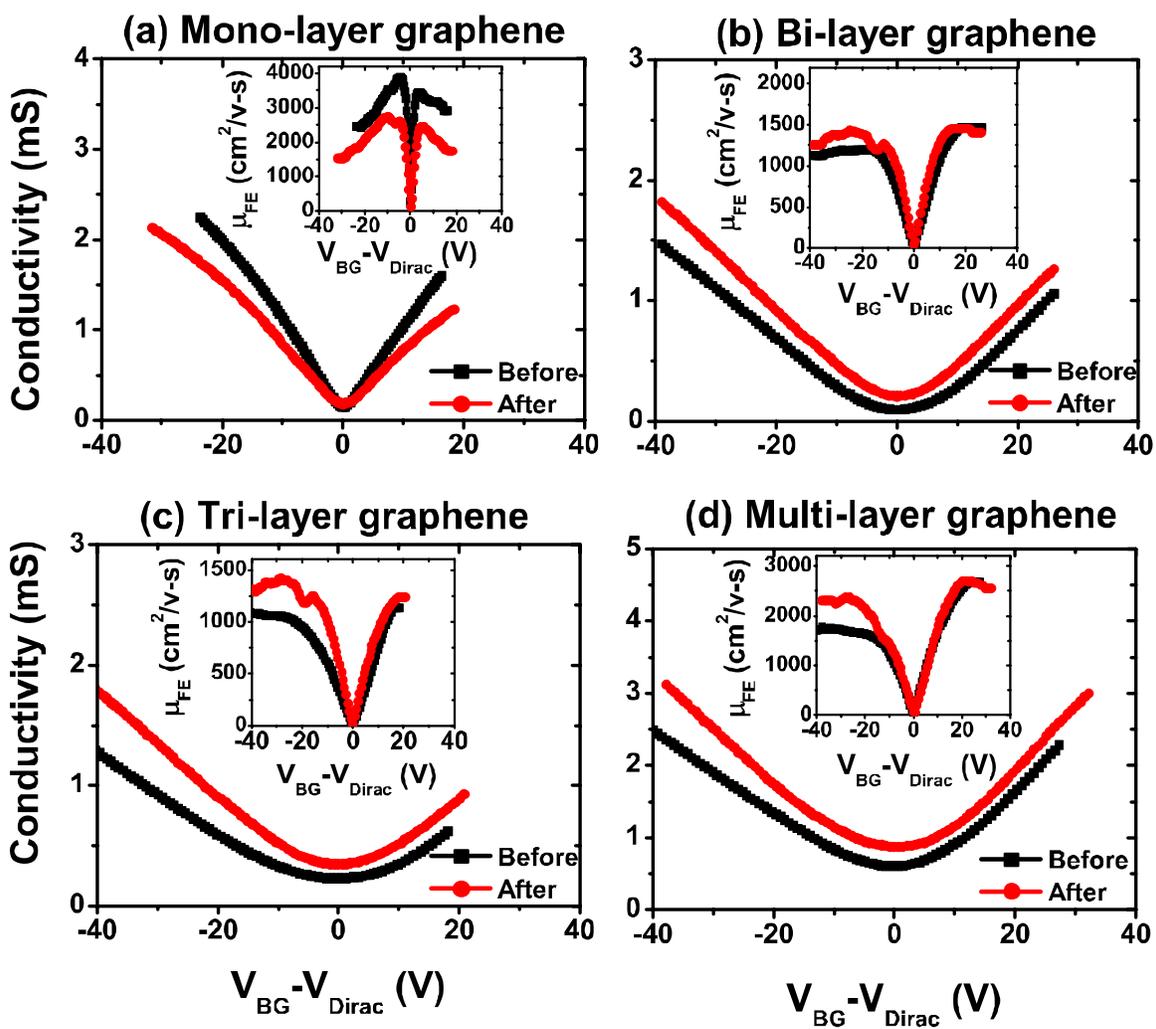

Figure 3.

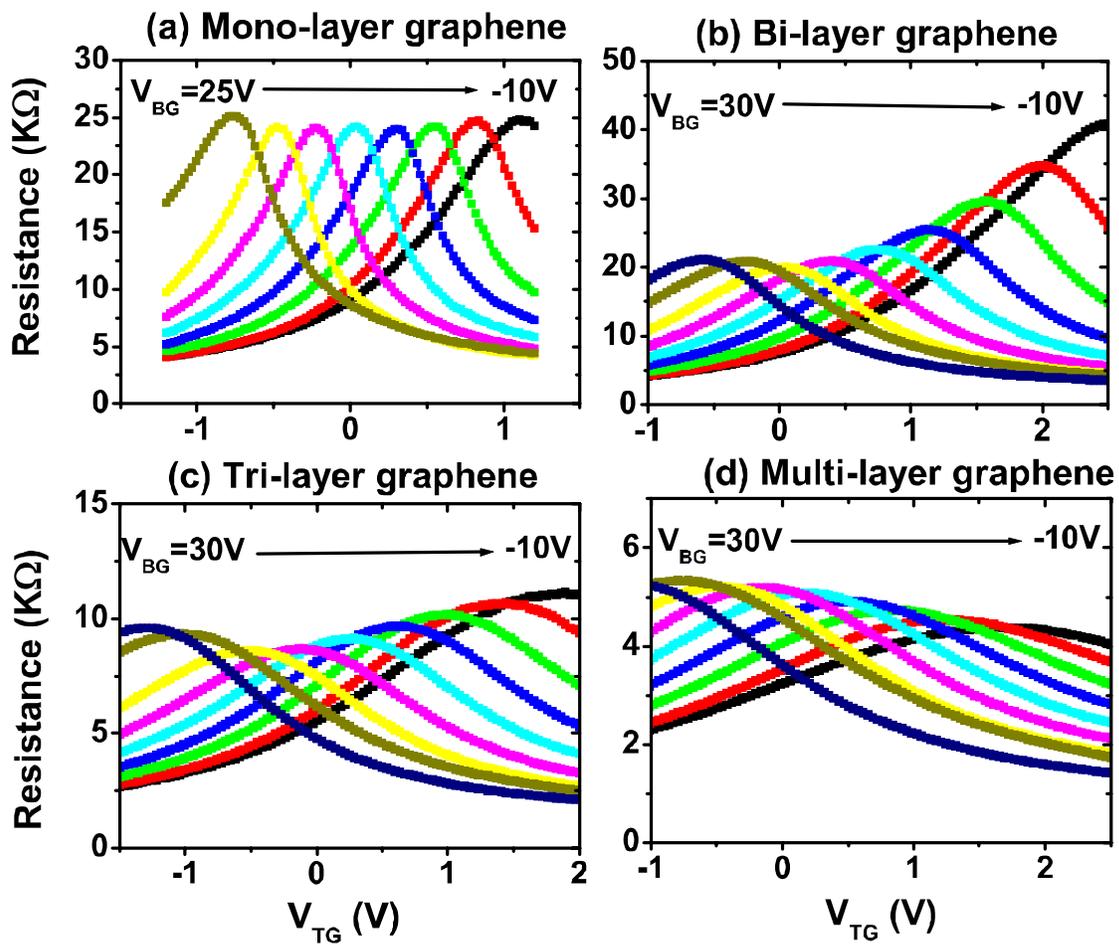

Figure 4.

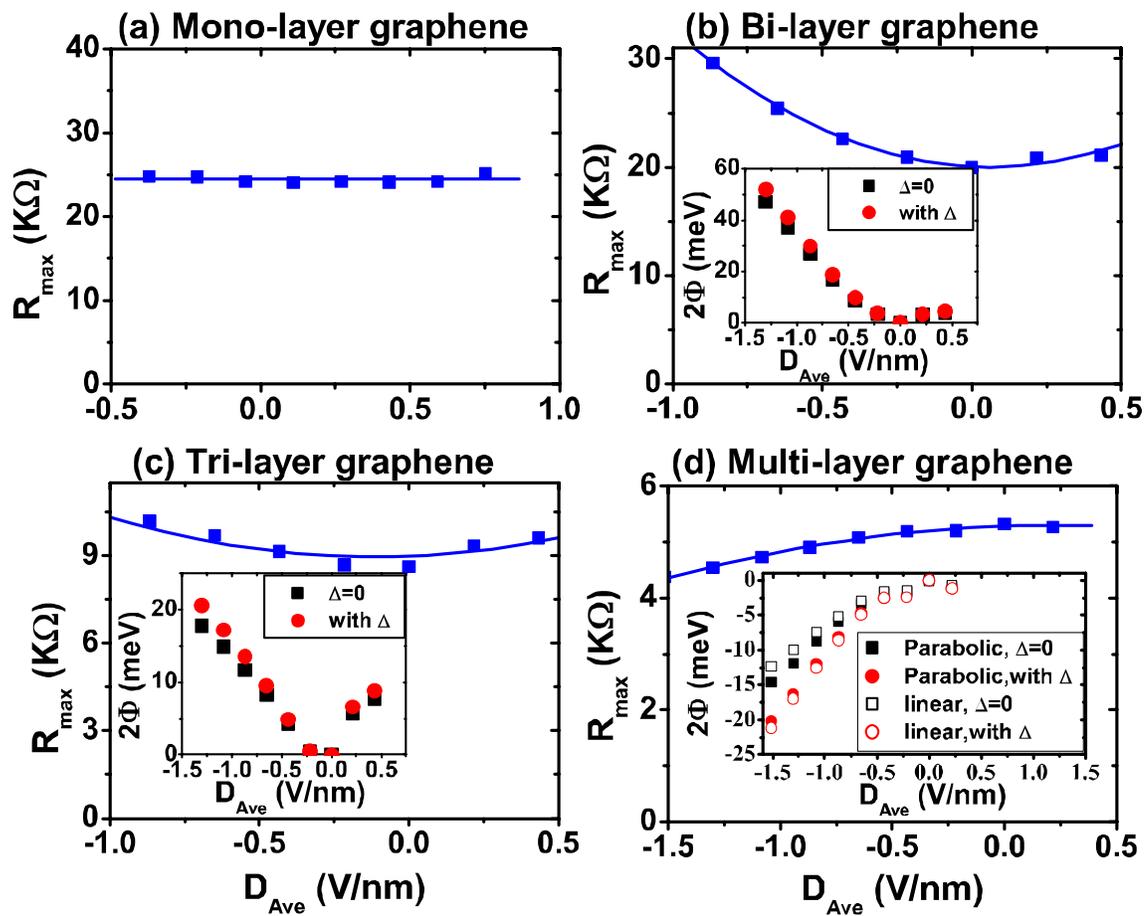

Figure 5.

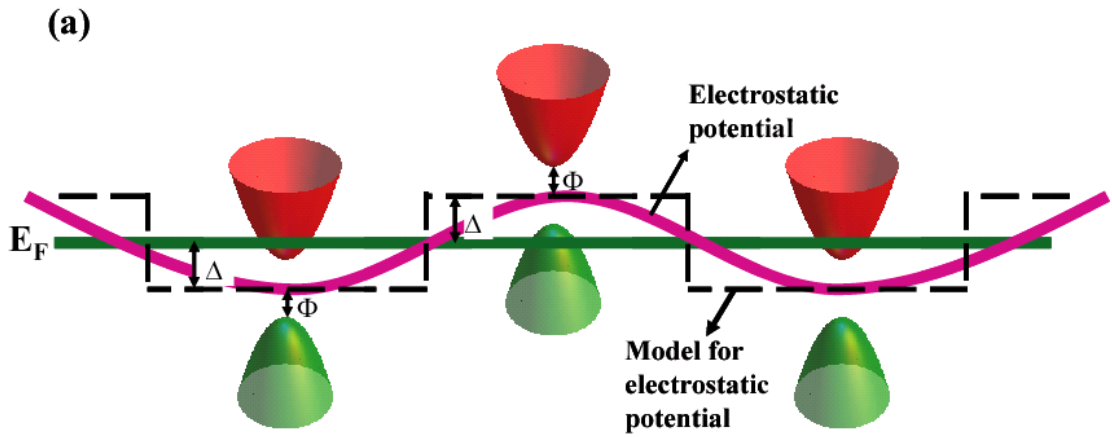

Figure 6.